\documentclass[prb,showpacs,twocolumn]{revtex4}
\usepackage{psfig}
\usepackage{graphicx}
\begin{document}
 
\title{Electronic structure and X-ray \\ 
magnetic circular dichroism of CrO$_2$}

\author{V. Kanchana and G. Vaitheeswaran}
\affiliation{Max-Planck-Institut f\"ur Festk\"orperforschung,
Heisenbergstrasse 1, 70569 Stuttgart, Germany}
\author{M. Alouani}
\affiliation{Institut de Physique et de Chemis des 
Mate\'riaux de Strasbourg (IPCMS), 23 rue du Loess, 
67037 Strasbourg Cedex, France} 

\date{\today}
\begin{abstract}
A detailed theoretical study on the 
electronic structure and 
magnetic properties of half-metallic
ferromagnet CrO$_2$ was carried 
out by means of relativistic full-potential
linear muffin-tin orbital method 
within the generalized gradient 
approximation
(GGA) to the exchange correlation 
potential. Our calculation 
favours the [001]
magnetization axis to be 
the easy axis of 
magnetization when compared to the [100] 
axis which is 
in agreement with the 
experiments. The calculated 
spin and
orbital magnetic
 moments of 
Cr agrees well with the 
experimental and other
theoretical works. The 
Cr L$_{2,3}$ x-ray 
absorption and 
x-ray magnetic 
circular
dichroism (XMCD) spectra were 
calculated for both the quantization 
axis and compared with the experiment. In 
addition the oxygen K edge and 
XMCD spectra were
also calculated which compares 
well with the experiments. The 
XMCD sum rules
were used to compute the spin 
and orbital magnetic moments and results agree
quite well with the direct calculation.  
\end{abstract} \pacs{75.50.Ss, 71.20.Lp, 75.30.Et}

\maketitle

\section{Introduction}
                                 
Chromium dioxide is the only stochiometric binary oxide that is a ferromagnetic
metal and also is the simplest and best studied half metal \cite{Coey}.  In
recent years, magneto electronic devices such as spin valve field sensors and
magnetic random access memories have emerged, where both charge and spin of
electrons are exploited using spin-polarized currents and spin-dependent
conduction \cite{Prinz,Daughton}. The performance of such magneto-electronic
devices depends critically on the substantial spin polarization of the
ferromagnetic components.  CrO$_2$ attracts specific interest being the only
compound reported so far experimentally to possess 100\% spin polarization.
The half-metallic band structure of CrO$_2$ was confirmed by several
experimental techniques such as Andreev reflection, super conducting tunelling,
photoemission, point-contact magnetoresistance, x-ray absorption, resonant
scattering and Raman spectroscopy\cite{Ji,
Parker,Schmitt,Dedkov,Coey1,Coey2,Huang1,Gupta,Kurmaev,Iliev}.

Half-metallic ferromagnetic materials appear as potential candidates and a lot
of work is under progress to synthesis magnetic oxides such as CrO$_2$,
Fe$_3$O$_4$ and Sr$_2$FeMoO$_6$ which are found to posses high Curie
temperature (T$_c$). Recently Sr$_2$CrReO$_6$ a Cr based double perovskite is
found to have a largest T$_c$ of about 635 K which is the
 highest among the
double perovskite family\cite{kato2002}.  These 
compounds serve 
as perspective
materials in the field of
 spintronics\cite{Haghiri}. Numerous 
theoretical works
are available for CrO$_2$ explaining the electronic structure, bonding, optical
and magneto-optical properties and magneto-crystalline
anisotropy(MCA) etc\cite{Schwarz,Sorantin,Uspenskii,Lewis,
Korotin,Mazin,Guo,Oppeneer,
komelj,Toropova}.
Because of the uniaxial crystal 
structure, CrO$_2$ is expected 
to have a large
magnetic anisotropy which 
makes it the favoured material 
for magneto-optical
recording. Recently it has been shown that
the low-temperature experimental data were reproduced 
well by LSDA itself without taking into account the Hubbard 
$U$ correction confirming that the ordered phase of CrO$_2$ is
weakly correlated\cite{Toropova}.

The circular dichroism-type spectroscopy became a powerful tool in the study of the
electronic structure of magnetic materials\cite{Bagus,Ebert}.  It has been
recently demonstrated by Weller et al\cite{Weller} that x-ray magnetic circular
dichroism (XMCD) is also a suitable technique to probe MCA at an atomic scale,
via the determination of anisotropy of the orbital magnetic moment on a
specific shell and site. The x-ray absorption spectroscopy (XAS) using
polarized radiation probes element specific magnetic properties of compounds
via the XMCD by applying the sum rules to the experimental 
spectra\cite{Thole,Carra,Laan}.  The application of these sum rules to
itinerant systems, in particular to low symmetry systems, is debated since the
sum rules are derived from atomic theory\cite{Carra,Chen,Wu}.
Recently angle resolved XMCD 
technique has been applied by 
Georing et al\cite{Georing,Georing1,Gold} for 
a wide range of temperature
to investigate the 
anisotropies of orbital moments $l_z$ and magnetic 
dipole term $t_z$ of Cr atom in  
epitaxial CrO$_2$ films on TiO$_2$ substrate. 
From their XMCD studies they found 
large anisotropies of $l_z$ and $t_z$, where the latter 
was derived from the semi emprical van der Laan's method 
of moment analysis\cite{Laan2}.

Theoretical understanding of XMCD for magnetic 
material is not an easy task, and
several {\it ab initio} calculations have attempted to compute XMCD of
transition metals and rare-earth compounds\cite{Wu,Ebert1,Alouani,
Brouder,Guo2,Ankudinov,Galanakis1,Galanakis2}.  The {\it $L_2$} and {\it $L_3$}
edges involving electronic excitations of the $2p$-core electrons towards 
$d$-conduction states have attracted much attention due to the dependence of
the XMCD  spectra on the exchange-splitting and the spin-orbit coupling of both
initial core and final conduction states. Brouder and co-workers\cite{Brouder},
Guo\cite{Guo2}, and Ankudinov and Rehr\cite{Ankudinov} used multiple-scattering
theory to study XMCD but their method, although successful, has been applied to
systems with few atoms per unit cell, as their formalism is computationally
involved. Finally, atomic calculations, using crystal-field symmetry, were
widely applied to fit the experimental {\it $M_4,_5$} edges of the rare earths
and actinide compounds and the {\it $L_2,_3$} edges of early transition metals.
Because of the large number of parameters to fit, it is difficult to apply this
formalism to delocalized $3d$ states\cite{van}.  Though there are lot of
theoretical studies available explaining the electronic and magnetic properties
of CrO$_2$ theoretical studies on the XMCD spectra were not available.
Parallel to our study, Baadji and coworkers\cite{baadji} are investigating the
effect of Hubbard interaction on the magnetic properties and XMCD specta of
CrO$_2$ using the linear Augmented plane wave method.

In the present study, efforts 
were taken to study theoretically the L$_{2,3}$
edge of Cr and K-edge of oxygen of CrO$_2$ using relativistic full-potential
linear muffin-tin orbital method (FP-LMTO) method. In Section II we briefly
present the theoretical method used to calculate the electronic structure,
orbital magnetic moment and XMCD spectra. The density of states and magnetic
moments are discussed in Section III. In Section IV we calculate the XAS and
XMCD spectra of CrO$_2$ and compare with the experiment. 
The conclusion of the results are presented in Section V.

\section{Theoretical Details}
In the present work electronic structure calculations were performed using the
all-electron full-potential linear muffin-tin orbital (FP-LMTO)
method\cite{wills} including the spin-orbit interaction directly in the
Hamiltonian.  The exchange correlation potential is parametrized using the
generalized gradient approximation (GGA)\cite{pbe}.  In this method, space is
divided into non-overlapping muffin-tin spheres surrounding the atoms, and an
interstitial region.  Most importantly, this method assumes no shape
approximation of the potential, wave functions, or charge density.  The
spherical-harmonic expansion of the potential was performed up to $l_{max}=6$,
and we used a double basis so that each orbital is described using two
different kinetic energies in the interstitial region.  The basis set consisted
of the Cr ($4s$ $4p$ $3d$), O  ($2s$ $2p$) LMTOs.  We performed our
calculations using the experimentally determined structure and atomic
positions, i.e., the rutile structure with space group symmetry $P4_2/mmm$ and
with cell parameters $a=b=4.419$~\AA, $c=2.912$\,\AA \cite{Porta}. The radii of
the muffin-tin spheres used for Cr and O were 2.0 and 1.5 Bohr units
respectively.  To find the easy magnetization axis we calculated total energy
only  for [001] as well as [100] magnetization axis.
The integration in reciprocal space was 
performed using the tetrahedron
method\cite{jepsen72} and  1100 irreducible {\bf k} points in the Brillouin
zone (BZ), for the [001] magnetization axis, whereas for [100] magnetization
axis 1200 {\bf k} points were used using
the same BZ grid as for the previous
quantization axis.  To avoid numerical errors 
one has to use the same {\bf k}
grid for both spin quantization axis and let the symmetry 
of the crystal in
presence of the spin-orbit decide for the number of of irreducible {\bf k}
points. The theoretical XAS and XMCD spectra was 
calculated using the method described elsewhere\cite{Alouani}. 
This method was found to be successful in reproducing the 
experimental XAS and XMCD spectra of serval transition 
metal compounds\cite{Alouani,Galanakis1,Galanakis2}.

\section{Density of states and magnetic moments}
The electronic structure of  CrO$_2$ has been extensively reported in the
literature\cite{Schwarz,Sorantin,Uspenskii,Lewis,
Korotin,Mazin,Guo,Oppeneer,Toropova}
so in this section we briefly discuss the density of states and compare with
the ealier works.

 \begin{figure}
\begin{center}
\includegraphics[width=70mm,clip]{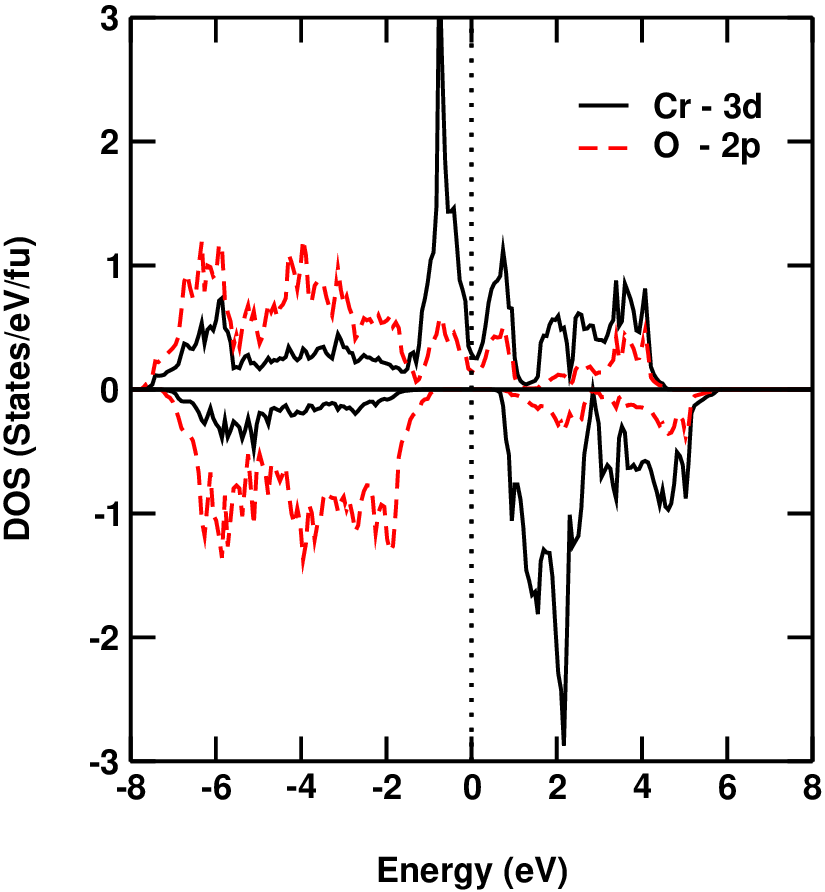}\\
\caption{(Color online) Calculated partial Cr-$3d$ and O-$2p$ density of states 
of CrO$_2$ the dotted line represents the Fermi level}
\end{center}
\label{figself6}
\end{figure}

Figure 1 shows the density of states (spin up and down) of Cr-$3d$ and O-$2p$
states. For the majority spin, the Fermi level lies near a local minimum of the
Cr 3d-$t_{2g}$ band with the DOS at the Fermi level. For the minority spin, the
Fermi level falls in a gap of 1.34 eV which is in agreement with the earlier
results\cite{Schwarz,Lewis,Korotin,Mazin,Guo}.  The calculated exchange
splitting between the majority and the minority spin main peaks of the Cr
3d-$t_{2g}$ band is found to be 2.3 eV. A similar splitting of 2.5 eV has been
found in FPLAPW calculations\cite{Mazin}, while a splitting of 2.3 eV
has been found in a recent FPLMTO calulations\cite{Guo}. All the exchange
splitting calculated within the LSDA or GGA are too small  when
compared with the measured large splitting 
of about 5 eV between 
the main peaks
in the occupied and the empty Cr-$3d$ DOS, from experimental photoemission
studies\cite{Tsujioka}.  It is interesting to note that 
half-metallic gap persists even with spin-orbit coupling. 
In general the spin-orbit coupling evantually 
destroys the half-metallic nature
in many Huesler alloys, dilute magnetic semiconductors,  
zinc blende type transition metal 
pnictides and double perovskites\cite{Mavropoulos,apl}.

The calculated spin and 
orbital magnetic moments 
along the [001] axis of magnetisation are found 
to be 1.99 $\mu_B$ and -0.045 $\mu_B$/atom for Cr and   
-0.08 $\mu_B$ and -0.0017 $\mu_B$/atom for O respectively.
The calculated spin magnetic moments of Cr and O are 
slightly higher when compared to the recent FP-LMTO calculations 
of Jeng and Guo\cite{Guo} within the LSDA. In the present 
calculation we have used 
GGA to the exchange correlation 
which will tend to increase 
the magnitude of the spin moment. When 
comparing the orbital 
magnetic moments of Cr and O they compares
well with the recent 
FLAPW calulations\cite{komelj}.
From the above results it can be 
seen that the the orbital magnetic moment of 
Cr is almost quenched and is also
antiparallel to the 
spin moment, which is 
consistent with Hund's rule for 3$d$
shells which are less 
than half-filled. A similar 
quenched orbital magnetic
moment for Cr has been recently
reported in the double perovskite
Sr$_2$CrReO$_6$\cite{apl}.  The orbital magnetic 
moment 
of the O atom is
parallel to the spin moment 
because the O-2$p$ shell is more than half-filled.
The calculated spin moment of 
oxygen is antiparallel to the spin moment of Cr
and hence the Cr and O are coupled 
antiferromagnetically.  
The calculated
orbital magnetic moments
of Cr and O are found to be -0.045 and -0.0017
$\mu_B$/atom respectively which are slightly lower
when compared with the experimental XMCD
measurments of  Huang et al in which they 
obtained a value of 
-0.06$\pm$0.02 $\mu_B$/Cr and -0.003$\pm$0.001
$\mu_B$/O respectively in the [001] axis of 
magnetization\cite{Huang2}.  While Georing et al\cite{Georing}
in their XMCD measurments on CrO$_2$ films 
found a relatively small contribution to the orbital magnetic 
moment of Cr in the [001] axis of magnetization 
which is found to be -0.02 $\mu_B$. When comparing 
the true spin 
moment of Cr atom Georing et al\cite{Georing} obtained a value of  
1.2 $\mu_B$. In order to explain the 
magnetic moment of 2 $\mu_B$ per unit cell which was obtained 
from the superconducting quantum intereference device (SQUID) 
measurments the above authors assumed a very large spin moment of 
about 0.4 $\mu_B$ per O which they try to explain interms of 
hybridization between the Cr and oxygen. Recently the same 
authors applied spin correction factor to their spectra from
which they were able to get a spin magnetic moment of 
2.4 $\mu_B$ which is slightly higher when compared to
the present theoretical and 
other experimental 
works\cite{Georing2}.
For the oxygen atom the spin
moments are once again too low when compared to the value estimated by
Georing et al\cite{Georing}. However it was recently shown by
Komelj et al\cite{komelj} that 
calculations with Hubbard $U$ = 3 eV yield a spin moment of 
0.1 $\mu_B$ which is a factor of 4 times smaller than the value 
obtained by Georing et al\cite{Georing}.
In order to find the easy axis of
magnetization we have
 done self consistent calculations only 
along the [001] and 
[100] magnetization axis. Our 
calculations find [001] axis to
be the easy axis of magnetization
which is in agreement with the 
experiments\cite{Spinu,Yang,Li,Miao,Gold}.

\section{X-Ray absorbtion and magnetic circular dichroism}
In this section we calculate, analyse and 
compare  the XMCD spectra with experiment.  
At the core level edge
XMCD is not only element specific but also 
orbital specific. For 3$d$
transition metals, the electronic states can be
 probed by the K, L$_{2,3}$,
M$_2,_3$ x-ray absorption and emission spectra. The 
dichroism at the L$_2$ and
L$_3$ edges is influenced by the spin-orbit 
coupling of the initial 2$p$ core
states. The large SO splitting of the 
core levels gives rise to a very pronounced 
dichroism in comparison with the
dichroism at the K-edge. In figure 2, we 
present the XAS and XMCD spectra for
Cr atom in the [001] axis of magnetization. 
We convoluted our theoretical spectra using 
a Lorentzian and Gaussian width of 
0.5 eV. The Gaussian represents the 
experimental resolution while Lorentzian 
corresponds to the width of the core hole. 

The calculated total absorbtion spectra 
corresponding of the Cr L$_{2,3}$ edge
was shown in the upper part(figure 2).  We have scaled
our spectra in a way that the
experimental and theoretical L$_3$ peaks in the absorption spectra have the
same intensity. The energy difference between 
the L$_3$ and L$_2$ peaks is given by the
spin-orbit splitting of $p_{1/2}$ and $p_{3/2}$ core states which 
is 8.65 eV. The
calculated value 
agrees well with the 
experimental value of 8.2-8.6
eV\cite{Georing2}. As far the absorption 
spectra is concerned the present theory 
represents well the experimental features but 
underestimates the intensity of 
the L$_2$ peak.

\begin{table}[tb]
 \caption{Spin and orbital magnetic moments as determined 
 from sum rules(SR) and self-consistently (SC) of 
 Cr-$d$ along the [001] axis of magnetization along with the 
 experimental values $^a$Reference\cite{Huang2},
 $^b$Reference\cite{Georing},$^c$Reference\cite{Georing2}}.
 \begin{ruledtabular}
 \begin{tabular}{cccccc}
  \multicolumn{2}{c}{$\mu_s$} &     \multicolumn{2}{c}{$\mu_l$}  & \multicolumn{2}{c}{Expt.}   \\
   SR     &  SC & SR      & SC    &$\mu_s$ &$\mu_l$  \\ \hline 
  1.82   &  1.99 &-0.044 &-0.045  & 1.9$\pm$0.1$^a$& -0.06$\pm$0.02$^a$\\
  & & &  &  1.2$^b$&  -0.02$^b$ \\
  & & &  &  2.15$^c$ & 

 \end{tabular}
 \end{ruledtabular}
 \end{table}
 
The intensity of the L$_3$ edge in the x-ray absorption
spectra when compared to the L$_2$ peak is a bit high which 
is in agreement with the experiment. The calculated
XMCD spectra along the [001] axis of magnetization is shown in 
the lower part. The present caculations reproduce well the shape 
of the experimental spectra.

\begin{figure}
\begin{center}
\includegraphics[width=70mm,clip]{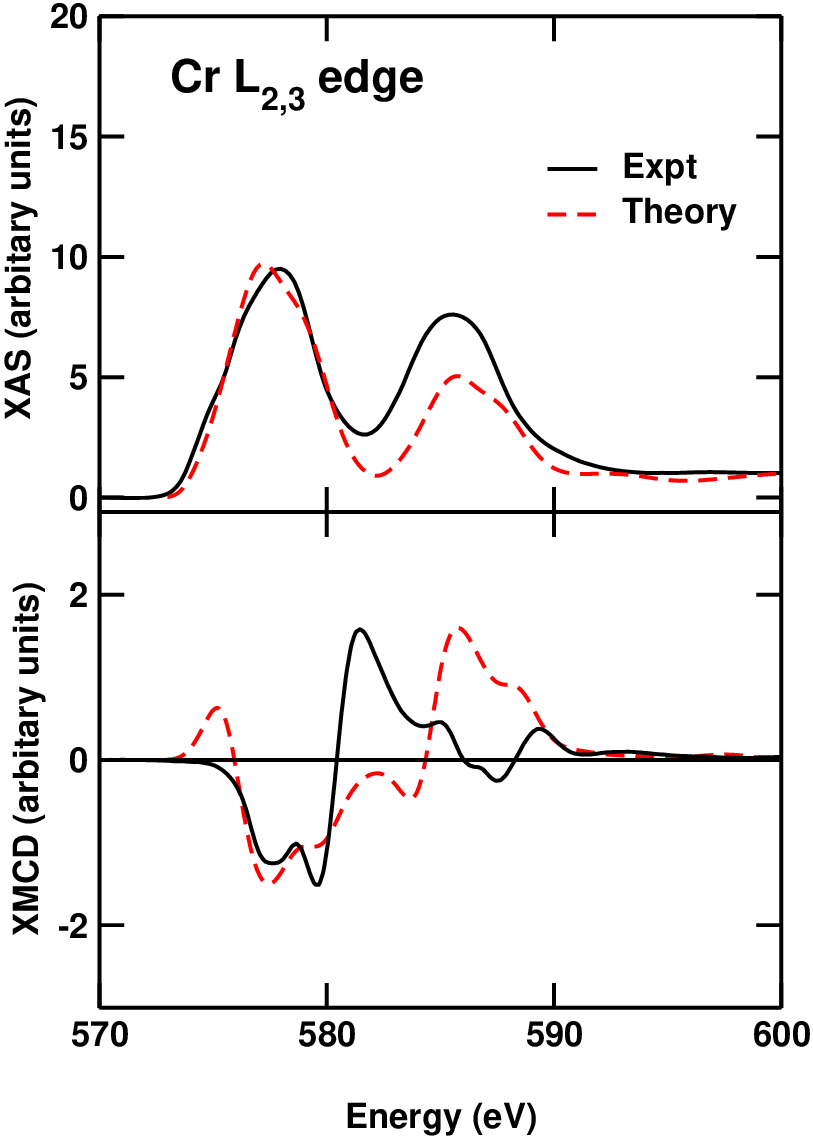}\\
\caption{(Color online) Cr L$_{2,3}$ XAS and XMCD spectra 
of  CrO2 along the [001] magnetisation
axis. The experimental spectra is taken from Ref.34}
\end{center}
\label{figself7}
\end{figure}

Using the XMCD sum rules we 
calculate the spin and orbital magnetic moment of
Cr-3$d$ state which 
are found to be 1.82 and -0.044 $\mu_B$/atom respectively and these 
values of spin and orbital magnetic moments 
and are in agreement with the values obtained 
from the self-consistent calculations. A deviation of about 
10\% in the magnetic moments when comparing the values 
from sum rules and direct calculation may result 
from the spectral overlap\cite{Georing3}. When comparing the spin 
and orbital magnetic moments obtained using the sum rules with 
the theoretical LSDA values of Huang et al\cite{Huang2} the spin 
moment of Cr is slight lower when compare to the theoretical
value of 1.9$\pm$0.1 $\mu_B$ whereas the orbital magnetic 
is found to be -0.06 $\pm$0.02 $\mu_B$ which 
is in good agreement with the present calculations. However the 
experimental studies of Huang et al\cite{Huang2} could not
provide quantitative information on the spin moment of Cr, 
because they cannot uniquely define which part of spectra belongs 
to the L$_3$ or L$_2$ edge.
When
comparing our magnetic moments obtianed from
sum rules with that of the XMCD measurments
of Georing\cite{Georing} the spin magnetic moment 
are too low
(1.2 $\mu_B$), whereas the spin magnetic moment(2.4 $\mu_B$) 
obtianed by the same authors after spin correction 
applied to the spectra\cite{Georing2} is slightly higher when compared to   
the present calculations. The 
orbital magnetic moments are also slightly higher (-0.044 $\mu_B$)
when compared to the experimental results of Georing in which they 
obtain a value of -0.02 $\mu_B$.

In figure 3, we have shown the oxygen K 
edge XAS and XMCD spectra along the [001]
axis of magnetization. 

\begin{figure}
\begin{center}
\includegraphics[width=70mm,clip]{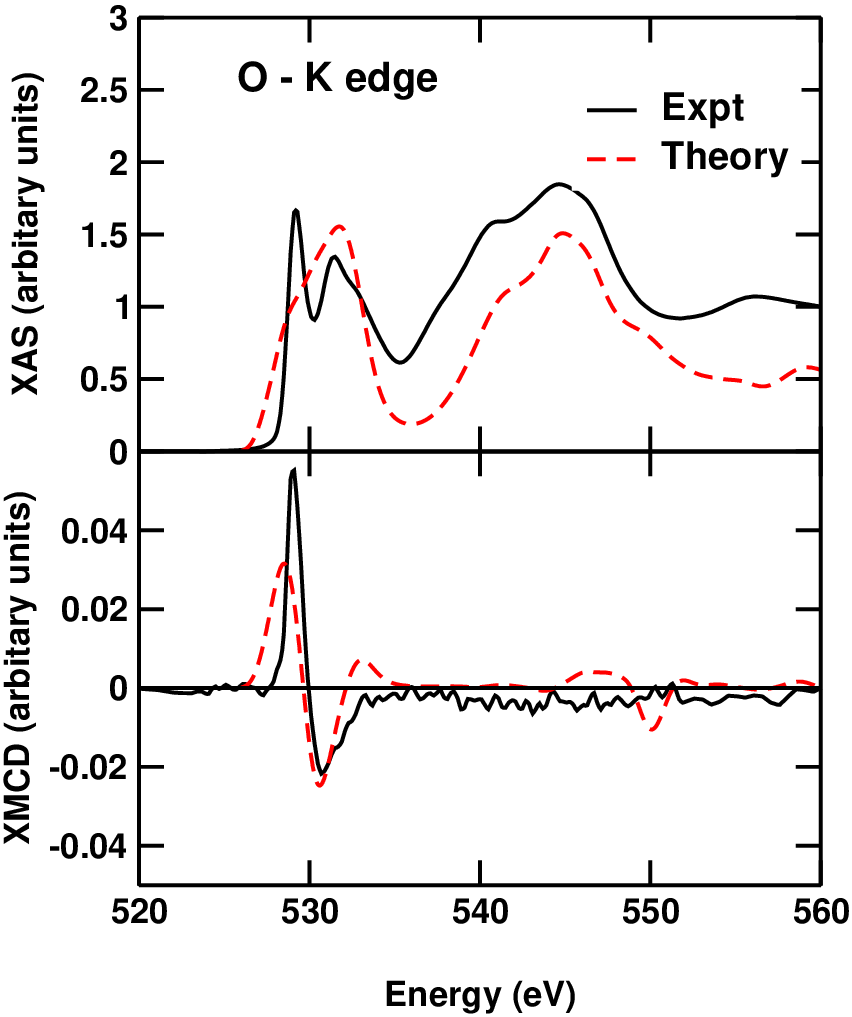}\\
\caption{(Color online) O K-edge XAS and XMCD spectra 
of  CrO2 along the [001] magnetisation
axis. The experimental spectra is teken from Ref.35}
\end{center}
\label{figself8}
\end{figure}

We used a similar 
broadening as that of the Cr L$_{2,3}$
for the oxygen K edge.  The oxygen 
K edge spectrum mainly comes from excitation 
of the 1$s$ state and is less intense 
when compared to the L$_{2,3}$ edge of Cr. The exchange
splitting  of the initial 1$s$ core state is 
extrmely small value, therefore, only
the exchange and spin-orbit
 splitting of the final 2$p$ state is responsible
for the observed dichroism at the K-edge. However the 
present calculation reproduce fairly well the experimentally 
observed O-K edge 
and XMCD spectra.

In figure 4, the Cr L$_{2,3}$ XAS and XMCD 
spectra of CrO$_2$ along the [100]
axis of magnetization is also shown. 

\begin{figure}
\begin{center}
\includegraphics[width=70mm,clip]{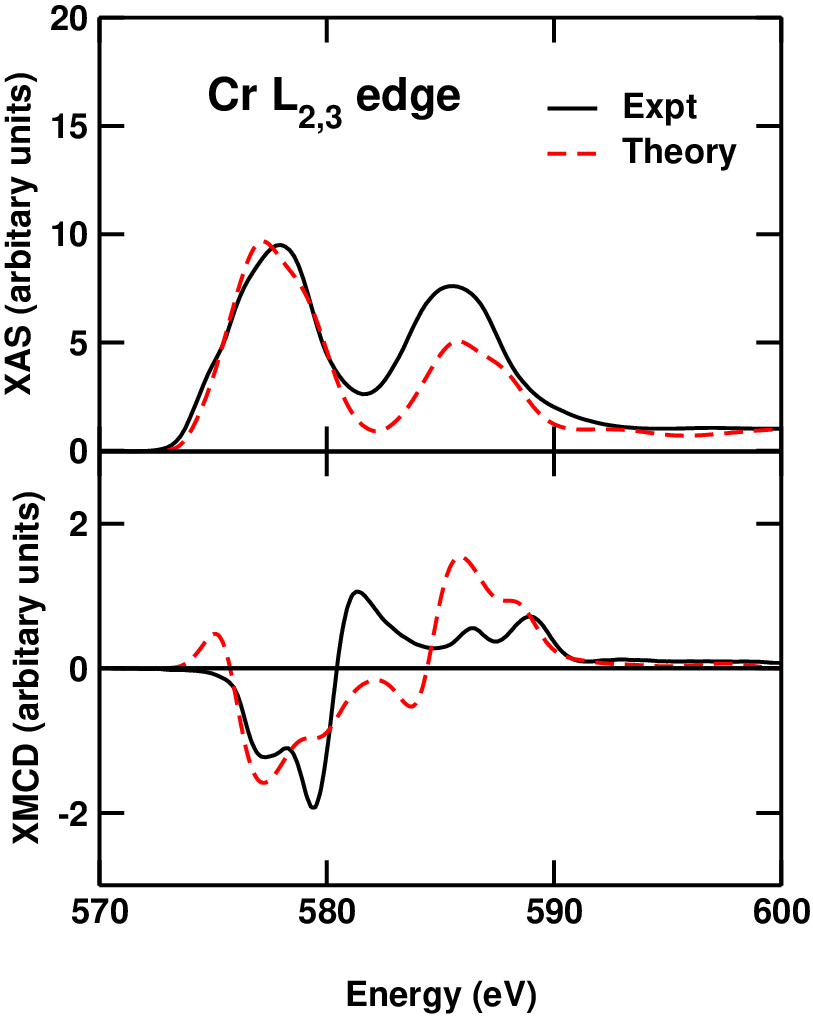}\\
\caption{(Color online) Cr L$_{2,3}$ XAS and XMCD spectra 
of  CrO2 along the [100] magnetisation
axis. The experimental spectra is taken from Ref.34}
\end{center}
\label{figself9}
\end{figure}

The 
core level splitting of 8.6 eV is
found in this axis which is in agreement with experimental 
value of 8.2-8.6\cite{Georing2}.
The theoretical XAS spectra agrees quite well with the 
experiments except the intensity of the L$_2$ peak is lower
when compared to the experiment. The calculated XMCD spectra
is shown in the lower part reproduce well the shape of the 
experimentally observed features. 
The calculated
XMCD intensity is relatively less in 
the[100] magnetization axis when compared
to the [001] axis of magnetization. For both the axes, the low
intensity of the
calculated L$_2$ edge makes the theoretical XMCD integrated L$_3$/L$_2$
branching ratio to be much larger than the experimental one. This discrepancy
may be due to the negligence of 3$d$ core-hole interaction.
Using the XMCD sum rules the spin 
and orbital magnetic moments of Cr atom along the [100] magnetization
axis is found to be 1.844 $\mu_B$ and -0.004 $\mu_B$, respectively. The 
caluculated spin magnetic moment agrees well with 
the earlier calculations\cite{Guo,komelj}.  The orbital magnetic 
moment obtained from the sum rules 
in the [100] axis is much lower when compares to the direct 
calculations. A similar discrepancy was observed in NiMnSb in 
which the orbital magnetic moments of Mn obtained from sum rules
are much lower when compared to the direct 
calculation\cite{Galanakis1}.However the theoretical orbital 
magnetic moments are much lower when compared to the experimental 
values of -0.09 $\mu_B$\cite{Georing}. 

The oxygen orbital magnetic moment is almost quenched in the
[100] axis, which are similar to the earlier theoretical 
works\cite{Guo,komelj} resulting in a weak intensity of 
the oxygen K edge in
contrast to the experiments, which show the same 
angular dependency for the Cr-3$d$ orbital 
projections and the O-K edge XMCD.

\section{Conclusions}
In the present work we have carried out a detailed theoretical study on the
magnetic properties of CrO$_2$. The calculated half-metallic band structure of
CrO$_2$ is in agreement with earlier studies. Our calculation confirms
the c-axis to be the
easy axis of magnetisation. The calculated XAS and XMCD
spectra of Cr L$_{2,3}$ and oxygen K edge compares 
fairly well with the experimental
results. The L$_3$/L$_2$ branching ratio is higher when compared to the
experiments which could be improved by including the 3$d$
core-hole interaction.  In addition the spin and orbital magnetic moments
obtianed from the sum rules are compared with the direct calculations.

\acknowledgments
The authors acknowledge Dr. E. Goering for sharing experimental details. 
M.A. thanks  O.K. Andersen for an invitation to the MPI during
the completion of this work.


 \end{document}